\documentclass[11pt]{article}
\usepackage{a4wide}
\usepackage{amsfonts,amsmath}
\usepackage{psfrag}
\usepackage{graphics}
\usepackage{subfigure}
\usepackage{ifpdf}
\usepackage{epsf}
\usepackage{color}
\ifpdf
\usepackage{epstopdf}
\usepackage{hyperref}
\else
\usepackage[hypertex]{hyperref}
\fi

\newcommand{\vev}[1]{{\langle\;{#1}\;\rangle}} 
\newcommand{\eref}[1]{(\ref{#1})}

\newcommand{\blank}[1]{}

\newcommand\sect[1]{\section{#1}\setcounter{equation}0} 

\newcommand\void[1]       {}

\newcommand\be            {\begin{equation}}
\newcommand\bea           {\begin{eqnarray}}
\newcommand\rd             {{\mathrm d}}
\newcommand\ee            {\end{equation}}
\newcommand\eea           {\end{eqnarray}}

\def\Bone{\hbox{1\!\!\!\!1}}

\renewcommand\vec[1]{{\vert{#1}\rangle}}

\newcommand\cev[1]{{\langle{#1}\vert}}

\newcommand\vac{{\vec 0}}

\def\3pt#1#2#3{{\langle{#1}|{#2}|{#3}\rangle}}

\def\cbI#1#2#3#4#5#6#7#8{
\setlength{\unitlength}{#1sp}%
\hbox to 2.8cm{\raise -2mm
\vbox{
\begin{picture}(1990,757)(4600,-2483)
\thicklines
{\put(4850,-1860){\line( 1,0){800}}}%
{\put(4850,-2460){\line( 1,0){800}}}%
{\put(5250,-1860){\line( 0,-1){600}}}%
\put(4775,-2460){\makebox(0,0)[rc]{$#2$}}
\put(4775,-1860){\makebox(0,0)[rc]{$#3$}}
\put(5750,-1860){\makebox(0,0)[lc]{$#4$}}
\put(5750,-2460){\makebox(0,0)[lc]{$#5$}}
\put(5300,-2160){\makebox(0,0)[lc]{$#6$}}
\end{picture}}}%
}

\def\cbb#1#2#3#4#5#6#7#8{
\setlength{\unitlength}{#1sp}%
\hbox to 3.6cm{\raise -2mm
\vbox{
\begin{picture}(2700,800)(4301,-2463)
\thicklines
{\put(4800,-1960){\line( 0,-1){500}}}%
{\put(5700,-1960){\line( 0,-1){500}}}%
{\put(4250,-2460){\line( 1, 0){2000}}}%
\put(4200,-2460){\makebox(0,0)[rc]{$#2$}}
\put(4800,-1900){\makebox(0,0)[cb]{$#3$}}
\put(4800,-2560){\makebox(0,0)[ct]{$#7$}}
\put(5250,-2380){\makebox(0,0)[cb]{$#4$}}
\put(5700,-1900){\makebox(0,0)[cb]{$#5$}}
\put(5700,-2560){\makebox(0,0)[ct]{$#8$}}
\put(6350,-2460){\makebox(0,0)[lc]{$#6$}}
\end{picture}}%
}}

\def\thefootnote{\fnsymbol{footnote}}
\begin{document}

\begin{flushright}  {~} \\[-12mm]
{\sf KCL-MTH-11-12}\\[1mm]
\end{flushright} 

\thispagestyle{empty}

\begin{center} \vskip 14mm
{\Large\bf   
Moduli space coordinates and excited state $g$-functions}\\[20mm] 
{\large 
G.M.T.~Watts~~\footnote{Email: gerard.watts@kcl.ac.uk}
}
\\[8mm]
Department of Mathematics, King's College London,\\
Strand, London WC2R 2LS -- UK

\vskip 22mm
\end{center}

\begin{quote}{\bf Abstract}\\[1mm]
We consider the space of boundary conditions of
Virasoro minimal models formed from the composition of a collection of
flows generated by $\phi_{1,3}$ 
between conformal boundary conditions.
These have recently been shown to fall naturally into a sequence, each
term having a coordinate on it in terms of a boundary parameter, but
no global parameter has been proposed.
Here we investigate the idea that has been put forward that the
overlaps of particular bulk 
states with the boundary states give natural coordinates on the moduli
space of boundary conditions. 
We find formulae for these overlaps using the known
thermodynamic Bethe Ansatz descriptions of the ground and first
excited state on the cylinder and show that they give a 
global coordinate on the space of boundary conditions, showing it is
smooth and compact as expected.

\end{quote}

%\tableofcontents

\vfill
\newpage 

\setcounter{footnote}{0}
\def\thefootnote{\arabic{footnote}}

\sect{Introduction}

The space of boundary conditions for Virasoro minimal models have been
investigated quite extensively. 
Distinguished amongst these boundary conditions are 
the conformal boundary conditions which have been classified by Cardy
\cite{Cardy89} and
Petkova et al. \cite{BPPZ00}.
These can be connected by renormalisation group flows generated by
relevant boundary fields. 
A special class of these are the integrable boundary flows
generated by the field $\phi_{1,3}$. 
These flows were first considered by Recknagel et al \cite{RRS00}. They
can be joined into a sequence which appeared in
\cite{LSS98}\footnote{It should be noted that the actual equations for
the g-functions of the bulk off-critical flows in this paper are not
correct, but exactly at the bulk critical point they do correctly
reproduce the sequence of boundary flows} 
and has
arisen naturally in recent 
work by Gaberdiel et al. \cite{FGSC09} and Dorey et al. \cite{DTW10}.
In \cite{FGSC09}, the flows fall into separate pieces which then have to
be joined together by hand to form the whole sequence; in
\cite{LSS98,DTW10} there is a separate parameter which can take the bulk
theory slightly off-critical in which case the whole sequence of
boundary flows arises with a single parameter, but as the bulk theory
becomes critical (conformal) the boundary flows again break into
disjoint sets.
Finally, the same sequence arises when the truncated conformal space
approach is used 
to study boundary flows \cite{W2011}; again this sequence disappears
as the truncation level regulator is removed.

In this paper we investigate the idea suggested in \cite{DRTW00} that
the overlaps of bulk states with boundary states could provide useful
coordinates on the space of boundary conditions.
The basic such overlap is the ground state overlap which is also
called the g-function or ground state entropy.
For purely elastic scattering theories, there are general formulae for
mixed
bulk--boundary flows \cite{DFRT04,DLRT06} which reduce to those in 
\cite{LMSS95} for purely boundary flows (The results in \cite{LMSS95} are not
correct when there is a simultaneous bulk perturbation).
The minimal models have been investigated in
\cite{DTW10} using a relationship with the purely-elastic staircase
model but at present general results for kink models are not known.

We start with a consideration of the tri-critical Ising model (TCIM
for short).
In the approximate mean-field approach \cite{Giokas2011}, the natural
coordinate to take on the space of boundary conditions is the
magnetisation at the boundary. In the conformal case the bulk spin
field diverges as it approaches the boundary
but we can consider a similar quantity, the expectation of the
bulk spin field on a disk. The bulk spin field in the TCIM is
$\sigma=\varphi_{2,2}$ \cite{LCM1} and the expectation value on a disk
is the ratio of 
the overlaps  of the states $\vec{\sigma}$ and $\vac$ with the
boundary state.

The g-functions along the TCIM boundary flows were derived by
Nepomechie and Ahn in \cite{NA}\footnote{The g-functions derived in
\cite{NA} are only correct for the purely boundary flows; when bulk
perturbations are included they suffer
from the same problems described in \cite{DRTW00}}
as well as being 
implicitly defined in \cite{BLZ94}.  
Nepomechie and Ahn did not consider the excited state g-functions
corresponding to 
the state $\vec\sigma$, but we can take the bulk TBA equations for
this state proposed in \cite{f,KMB370} and put them into the formulae of
\cite{NA} we do indeed obtain functions which correctly
interpolate the overlap of $\vec\sigma$ with the conformal boundary
state; these are our proposals for the excited state g-functions in
this case. 

The paper is organised as follows.
In section \ref{sec:tcim} we summarise the known facts about the
tri-critical Ising model, its boundary conditions and related TBA
systems.
In section \ref{sec:tcimgfn} we present our proposals for the excited
state TBA g-functions and show that $\vev\sigma$ provides a good
coordinate on the space of boundary conditions.
In section \ref{sec:gen} we present the generalisation to an arbitrary
diagonal unitary minimal model.

\sect{The boundary tri-critical Ising model}
\label{sec:tcim}

The space of boundary conditions and their interpolating flows in the
TCIM was first presented by Affleck in \cite{Affleck00} and includes
the following
\be
 (-) \longleftarrow (-0) \longrightarrow (0)
  \longleftarrow (0+) \longrightarrow (+)
\;.
\label{eq:basicflow}
\ee
where the boundary conditions are labelled according to their
interpretation in terms of 
the allowed values of a lattice spin taking values in $\{-1,0,+1\}$;
Each flow is generated by a field $\phi_{1,3}$ of conformal
weight $3/5$. 
The boundary conditions can also be labelled by their Kac labels
\cite{YBk}: 			
there are three 
fixed boundary conditions
$(11)=({+}),(21)=(0), (31)=({-})$ and two variable boundary conditions
$(12)=(0{+})$ and $(13)=({-}0)$. The final elementary boundary state
$(22)=(d)$ is not part of this sequence.

We would like to give a parametrisation of the whole space
\eref{eq:basicflow} -- it appears to be an entirely natural sequence
of flows which form a one-dimensional subspace of the space of
boundary conditions and if the space of boundary conditions is a
genuine moduli space one ought to be able to put coordinates on it.

In terms of a mean-field approximation \cite{Giokas2011}, the space
\eref{eq:basicflow} can be 
parametrised by the value 
of the boundary magnetisation, $m$. This takes values $-1\leq m\leq1$
and in particular
\be
 m_{(-)} = -1
\;,\;\;
 m_{(-0)} = -0.4774
\;,\;\;
 m_{(0)} = 0
\;,\;\;
 m_{(0+)} = 0.4774
\;,\;\;
 m_{(+)} = 1
\;,
\ee
In the conformal field theory, the spin field is represented by the
conformal field $\sigma=\varphi_{2,2}$ of conformal weight $3/40$
\cite{LCM1}. This field 
diverges as it approaches any of the conformal boundary conditions
apart from $(0)$ and so it does not have a finite boundary
magnetisation. 

We can choose instead a related quantity, $b_\alpha^\sigma$, the
expectation value of the 
bulk spin on a unit disk with the appropriate boundary condition. 
This is directly related to the divergence of the spin as it
approaches the boundary, being just the bulk-boundary structure
constant.

If we restrict for now to the case of a conformal boundary condition
$\alpha$, we can define the boundary state
\be
  \cev\alpha = \sum_j g_\alpha^i\, \langle\langle j | 
\;,
\ee
where $\langle\langle j |$ are the Ishibashi states for the
representation $j$ 
The coefficient  $g_\alpha^1
\equiv g_\alpha$ is the g-function introduced in \cite{AL}.
The structure constant 
$b_\alpha^i$
appears in
the bulk-boundary operator product expansion of a bulk field
$\varphi_i$ approaching a conformal boundary condition $\alpha$ at
$y=0$ 
\be
  \varphi_i(x,y)
= b_\alpha^i
\;\Bone\;(2y)^{-x_i} + \ldots
\;.
\ee
This allows one to calculate the partition function on the unit disk
with the insertion of $\varphi_i$ in two ways as
\be
  \langle \alpha| \varphi_i(0) \vac
= \langle \alpha \vec i
= g_\alpha^i
\;,\;\;
  \langle \alpha| \varphi_i(0) \vac
= b_\alpha^i %{}^\alpha B_{i}^{\sBone} }
  \,{\vev \Bone }_\alpha
= b_\alpha^i %{}^\alpha B_{i}^{\sBone} }
  \,{\langle\alpha\vac}
= b_\alpha^i %{}^\alpha B_{i}^{\sBone} }
  \,g_{\alpha}
\;,
\ee
so that $b_\alpha^i$ %the bulk-boundary structure constant 
can be found as
the expectation value of the bulk field,
\be
 b_\alpha^i % \equiv {}^\alpha B_{i}^{\sBone} }
= \frac{g_{\alpha}^i}{g_\alpha}
= \frac{ \langle\alpha| \varphi_i(0) \vac}
       { \langle\alpha  \vac}
= \frac{ \vev{ \varphi_i(0) }_\alpha }{\vev{ \Bone }_\alpha }
\;.
\ee
For the diagonal unitary minimal models, $g_\alpha^i=
S_\alpha{}^i/\sqrt{S_1{}^i}$ where  
$S_{rr'}{}^{ss'}$ is the modular S--matrix  \cite{Cardy89}.
%$=(-1)^{1+rs'+r's}(8/m(m+1))^{1/2}\sin(rr'(m+1)/m)\sin(ss'm/(m+1)) $.
The numerical values of the $g_\alpha^i$ and some useful combinations
are in table \ref{tab:tcimgs}.
\begin{table}
\[
{\renewcommand{\arraystretch}{1.4}
\begin{array}{l||c|c|c|c|c}
\alpha & (-) & (-0) & (0) & (0+) & (+) \\
\hline
\hline
g_\alpha 
& 0.5127 & 0.8296 & 0.7251 & 0.8296 & 0.5127
  \\
\hline
g_\alpha^{\sigma} 
& -0.7756 & -0.4793 & 0 & 0.4793 & 0.7756
  \\ \hline
b_\alpha^\sigma %=g_\alpha^\sigma/g_\alpha
&-1.513
&-0.578
&0
&0.578& 1.513
\\\hline
g_\alpha/g_{(+)}
&1. &1.618&1.414&1.618&1.\\
\hline
\log|g_\alpha/g_{(+)}|
&0. &0.4812&0.3466&0.4812&0.\\
\hline
b_\alpha^\sigma/b_{(+)}^{\sigma}
&-1.&-0.618&0&0.618&1.
\\\hline
\log|b_\alpha^\sigma/b_{(+)}^{\sigma}|
&0.&-0.4812&-\infty&-0.4823&0.
\end{array}
}
\]
\caption{Numerical values for some constants in the tri-critical Ising
  model}
\label{tab:tcimgs}
\end{table}

We would like to take $b_\alpha^\sigma$ as the coordinate
along the space of flows \eref{eq:basicflow}. As can be seen in table
\ref{tab:tcimgs}, this increases along the fixed points of the flow
\eref{eq:basicflow} and we expect this will remain true for all the
flow, not just the fixed points, but to check that we need to find a
way to calculate it at non-critical points. 
To do that we will use TBA methods which are explained in the next
section.

\sect{The TBA description of the tri-critical Ising model}
\label{sec:tcimtba}

We want to find equations for $b_\alpha^i=g_\alpha^\sigma/g_\alpha$
when $\alpha$ is a
perturbed boundary condition. 
%This is the ratio of $g_\alpha^\sigma$
%and $g_\alpha$. The latter is 
The standard $g$-function $g_\alpha$ can be
found by integrating suitable functions from the bulk TBA system
against kernels determined by the boundary reflection matrices;
$g_\alpha^\sigma$ is an ``excited $g$-function'' or an
overlap of the boundary state with an excited state which is expected 
to be given by integrating excited state TBA functions against the
same kernels. In the case of the Yang-Lee model the first excited
state can be found by analytic continuation of the ground state in the
perturbation parameter \cite{DT} and so the excited state g-functions can also
be found that way. For the TCIM, the first excited state is in a
different sector of the model (it is odd under the $Z_2$ symmetry rather than
even like the ground state) so it is very unlikely to be simple to
connect it to the ground state, but there are related TBA-like
equations which describe it, as proposed in \cite{KMB370}.

The TBA equations for the ground state come in three varieties: 
the massless TBA for
the model $M_{4,5} + \lambda\varphi_{1,3}$ given by the addition of the
bulk field $\varphi_{1,3}$ with a positive coupling which flows to the
critical Ising model \cite{Zamo9}; 
the massive TBA for
the model $M_{4,5} - \lambda\varphi_{1,3}$ given by the addition of the
bulk field $\varphi_{1,3}$ with a negative coupling which flows to a
model with  one massive kink \cite{Zamo9b}; 
the kink TBA which 
describes the purely critical model and can be found as the common limit
of the massive and massless TBA systems.
We shall restrict ourselves to the kink system which is all we need to
describe the boundary flows in the critical bulk model.

\subsection{The kink TBA equations for the ground state}

The kink TBA equations for the ground state are integral equations for
two functions $\epsilon_i(\theta)$,
\bea
      \epsilon_1(\theta) 
&=& - \int_{-\infty}^\infty \Phi(\theta-\theta')
      L(\epsilon_2(\theta'))
      \frac{\rd\theta'}{2\pi}
\nonumber\\
      \epsilon_2(\theta) 
&=&   \frac{r}{2}e^{\theta}
    - \int_{-\infty}^\infty \Phi(\theta-\theta')
      L(\epsilon_1(\theta'))
      \frac{\rd\theta'}{2\pi}
\;,
\label{eq:tba1}
\eea
where $\Phi(\theta)=\mathrm{sech}(\theta)$,
$L(\epsilon)=\log(1+e^{-\epsilon})$ and $r$ is an essentially
irrelevant parameter. 
The ground state energy $-c/12$ is given in terms of this system by
\be
 E = - \int_{-\infty}^\infty
       {r}e^{\theta} L(\epsilon_2(\theta)) 
       \frac{\rd\theta}{4\pi^2}
   = - \frac{7}{120}
\;.
\ee
The four separate flows in the sequence 
\eref{eq:basicflow} can each be labelled by a parameter $\theta_b$ 
for which the UV is given by $\theta_b\to-\infty$ and the IR by
$\theta_b\to+\infty$, and the g-functions are given as
\cite{NA,BLZ94}
\bea
    \log\big[ g_{(0+)\to(0)}(\theta_b) / g_{(+)} \big]
&=& \int_{-\infty}^\infty \Phi(\theta - \theta_b) L(\epsilon_1(\theta))
    \frac{\rd\theta}{2\pi}
\;,
\label{eq:gb1}
\\
    \log\big[ g_{(0+)\to(+)}(\theta_b) / g_{(+)} \big]
&=& \int_{-\infty}^\infty \Phi(\theta - \theta_b) L(\epsilon_2(\theta))
    \frac{\rd\theta}{2\pi}
\;.
\label{eq:gb2}
\eea

\subsection{The kink TBA equations for the first excited state}

The TBA equations proposed in \cite{KMB370} for the first excited
state $\vec\sigma$ of conformal weight $3/40$
are the very similar set
\bea
      \epsilon'_1(\theta) 
&=& - \int_{-\infty}^\infty \Phi(\theta-\theta')
      L'(\epsilon'_2(\theta'))
      \frac{\rd\theta'}{2\pi}
\nonumber\\
      \epsilon'_2(\theta) 
&=&   \frac{r}{2}e^{\theta}
    - \int_{-\infty}^\infty \Phi(\theta-\theta')
      L'(\epsilon'_1(\theta'))
      \frac{\rd\theta'}{2\pi}
\;,
\label{eq:tba2}
\eea
where 
$L'(\epsilon')=\log(1-e^{-\epsilon'})$. 
The energy $(2h_\sigma - c/12)$ of $\vec\sigma$ is given in terms of
this system as 
\be
 E = - \int_{-\infty}^\infty
       {r}e^{\theta} L'(\epsilon'_2(\theta)) 
       \frac{\rd\theta}{4\pi^2}
   = \frac{1}{60} 
\;.
\ee
If we substitute $L'(\epsilon_i)$ for $L(\epsilon_i)$ in equations
\eref{eq:gb1} and \eref{eq:gb2} we find we have constructed the TBA
formulae for the perturbed excited state g-functions:
\bea
    \log\big[ g^\sigma_{(0+)\to(0)}(\theta_b) / g^\sigma_{(+)} \big]
&=& \int_{-\infty}^\infty \Phi(\theta - \theta_b) L'(\epsilon'_1(\theta))
    \frac{\rd\theta}{2\pi}
\;,\\
    \log\big[ g^\sigma_{(0+)\to(+)}(\theta_b) / g^\sigma_{(+)} \big]
&=& \int_{-\infty}^\infty \Phi(\theta - \theta_b) L'(\epsilon'_2(\theta))
    \frac{\rd\theta}{2\pi}
\;,
\eea
and consequently we have formulae for the one-point function of the
spin field along the flows in terms of the solutions of the ground
state and excited state TBA equations:
\bea
    \log\big[ b^\sigma_{(0+)\to(0)}(\theta_b) / b^\sigma_{(+)} \big]
&=& \int_{-\infty}^\infty \Phi(\theta - \theta_b) 
   (L'(\epsilon'_1(\theta))-L(\epsilon_1(\theta)))
    \frac{\rd\theta}{2\pi}
\;,\\
    \log\big[ b^\sigma_{(0+)\to(+)}(\theta_b) / b^\sigma_{(+)} \big]
&=& \int_{-\infty}^\infty \Phi(\theta - \theta_b) 
    (L'(\epsilon'_2(\theta)) - L(\epsilon_2(\theta)))
    \frac{\rd\theta}{2\pi}
\;,
\eea
The g-functions and b-functions for the remaining two flows
$(-0)\to(0)$ and $(-0)\to(-)$ are found from these by the $Z_2$
spin-reversal symmetry.

\subsection{Results for the tri-critical Ising model}
\label{sec:tcimgfn}

\begin{figure}[htb]
\subfigure[The functions $\Phi*L(\epsilon_1)$. Also shown are the values of
  $\log g_\alpha$ for  $\alpha=(0), (0+)$ and $(+)$.
]{\scalebox{0.67}{\includegraphics{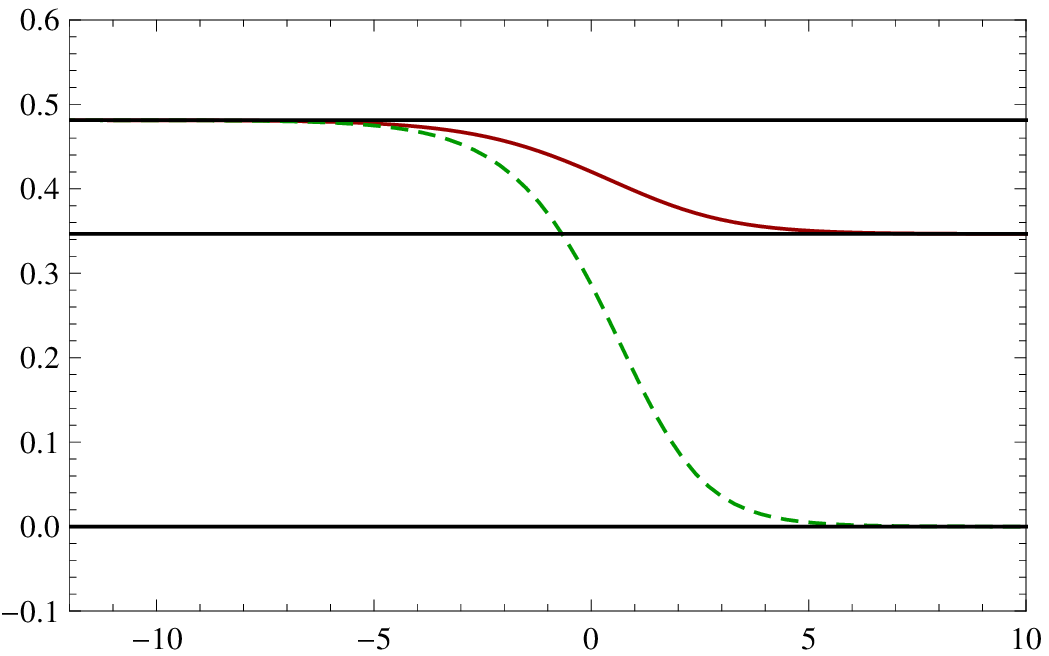}}\label{fig:m45klp}}
\hspace{5mm}
\subfigure[The functions $\Phi*L'(\epsilon'_1)$. Also shown are the values of
  $\log g_\alpha^\sigma$ for  $\alpha=(0+)$ and $(+)$.
]{\scalebox{0.67}{\includegraphics{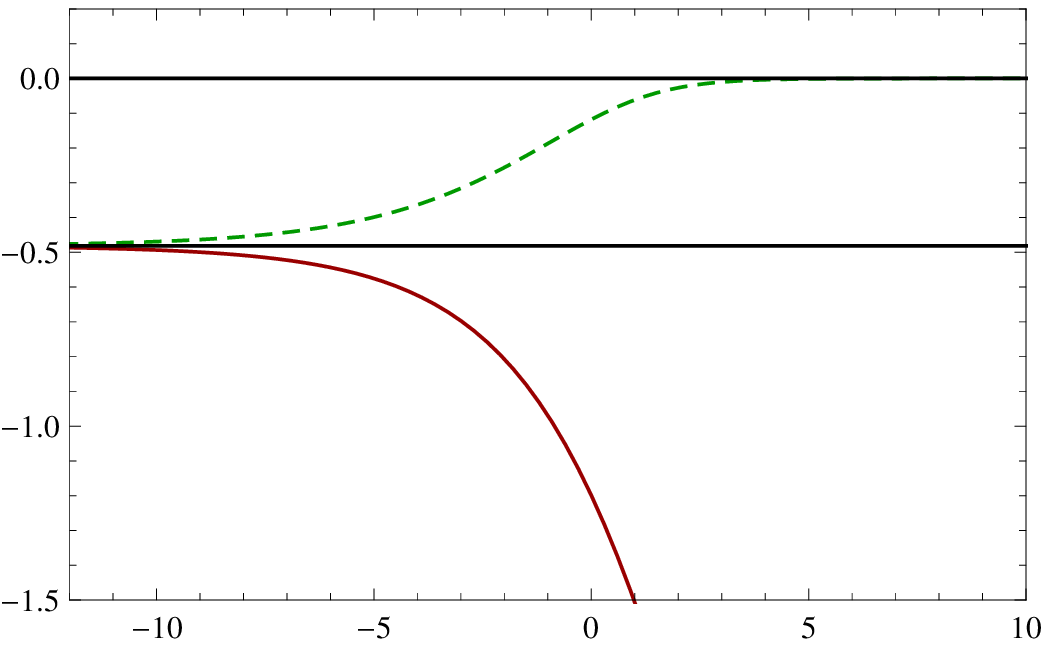}}\label{fig:m45kl2p}}
\caption{The functions $\Phi*L(\epsilon_i)(b)$ and $\Phi*L'(\epsilon'_i)(b)$
  for $M_{4,5}$ plotted against $b$ for $i=1$ (red, solid) and $i=2$
  (green, dashed). The   UV is on the left and the IR on the
  right. 
}
\label{fig:m45kls}
\end{figure}

The TBA equations \eref{eq:tba1} and \eref{eq:tba2} are easy to solve
by discretisation and iteration. 
In figure \ref{fig:m45klp} we plot the functions
$\Phi*L(\epsilon_i)(b)$ and in figure 
\ref{fig:m45kl2p} we plot $\Phi*L'(\epsilon'_i)(b)$, where the
convolution is 
$f*g(\theta)=\int f(\theta-\theta')g(\theta')\rd\theta'/(2\pi)$. These
can be seen to  
interpolate the UV values for large negative $b$ and the IR values for large
positive $b$.

\begin{figure}[htb]
\centerline{\scalebox{0.7}{\includegraphics{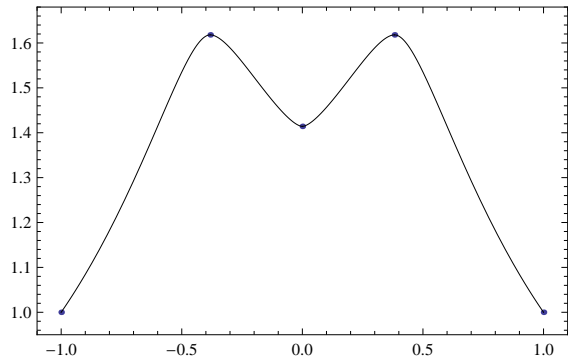}}}
\caption{%
A plot of 
$g_\alpha/g_{(+)}$ 
vs 
$b_\alpha^\sigma/b_{(+)}^\sigma$ 
for
the basic series of flows in the tri-critical Ising model. The dots
indicate the conformal boundary conditions, namely $(-)$, $(-0)$,
$(0)$, $(0+)$ and $(+)$ reading from left to right.
}
\label{fig:m45plot}
\end{figure}

It is also very helpful to plot the sequence of flows
\eref{eq:basicflow} in $\mathbb R^2$ with coordinates
$b_\alpha^\sigma/b_{(+)}^\sigma$ and $g_\alpha/g_{(+)}$ which we do in
figure \ref{fig:m45plot}.
This shows that $b_\alpha^\sigma$ provide a good global coordinate
along the sequence of flows, as well as highlighting the fixed points
which are the stationary points of $g_\alpha$.

\sect{The general minimal model moduli space}
\label{sec:gen}

We now propose the extension of the result in section \ref{sec:tcim}
to the general unitary minimal model $M_{m+2,m+3}$. 
The minimal model $M_{m+2,m+3}$ can be formulated in terms of a single spin
taking values in $\{ -1, \tfrac {2-m}{m}, \ldots \tfrac{m-2}{m},+1\}$ --
this can be thought of as the average value of the spins at each end
of an edge in the
standard RSOS realisation.
The spin field in the Landau-Ginzburg
description \cite{Zam86} related to the lattice realisation 
is again $\varphi_{22}$. There is a similar basic
sequence of boundary flows generated by the boundary field $\phi_{13}$:
given in terms of Kac labels, this is
\be
 (m{+}1,1) \longleftarrow 
 (1,m{+}1) \longrightarrow 
 (m,1) \longrightarrow
  \cdots
\longleftarrow (13) \longrightarrow (21)
\longleftarrow (12) \longrightarrow (11)
%\;.
\label{eq:basicflow2}
\ee
The $(r,1)$ boundary states correspond to fixed boundary spins
$\sigma_B=(m+2-2r)/m$ and the
$(1,r)$ boundary states correspond to spins which can take two
consecutive values. 

\subsection{The case of $m$ even: $M_{2p+2,2p+3}$}

\begin{figure}[htb]
\centerline{\scalebox{0.7}{\includegraphics{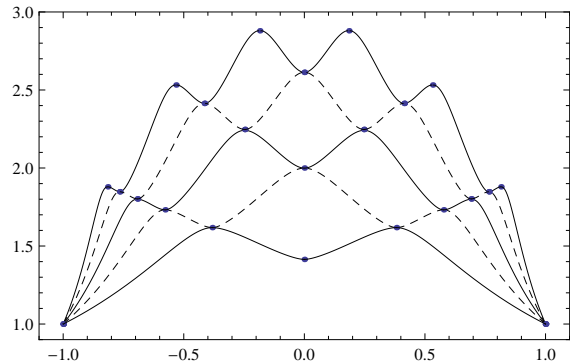}}}
\caption{%
A plot of $g_\alpha/g_{(+)}$ 
vs.
$b_\alpha^\sigma/b_{(+)}^\sigma$
for
the basic series
of flows \eref{eq:basicflow2} in the models $M_{r,r+1}$ for
$r=4,5,6,7$ and 8; the models with $r$ even are shown with solid
lines, those with $r$ odd by dashed.
The dots indicated the conformal fixed points.
}
\label{fig:m89plot}
\end{figure}

For the case of $m=2p$ even,
the ground state TBA equations
are given in \cite{Zamo9b} and the excited state TBA equations for the
state $\vec\sigma$ are again given in \cite{KMB370}.
The $g$-functions can be found in 
\cite{BLZ94} in terms of the eigenvalues of certain operators which
are simply related to the ground-state TBA functions; they are simple 
generalisation of those for the tri-critical Ising model.
The ground state kink TBA is a system of integral equations for $p$
functions $\epsilon_i$:
\be
  \epsilon_i 
= \delta_{i,m}\frac{r}{2}e^{x}
  - \frac{1}{2\pi}\sum_j I_{ij}\; \Phi * L(\epsilon_j) 
\;,
\label{eq:gfn1}
\ee
where $I_{ij}$ is the incidence matrix of the $A_m$ Dynkin diagram.
The first excited state kink TBA of \cite{KMB370} is the related set
\be
  \epsilon'_i 
= \delta_{i,m}\frac{r}{2}e^{x}
  - \frac{1}{2\pi}\sum_j I_{ij}\; \Phi * L_j(\epsilon'_j) 
\;,
\label{eq:gfn2}
\ee
where $L_j\equiv \log(1 + s_j e^{-\epsilon_j})$ and the sign $s_j=-1$
for $j=p,p+1$ and $s_j = 1$ otherwise. 
From \cite{BLZ94} we find that the g-functions for the flows in
\eref{eq:basicflow2} with $r=2,..,p+1 $ are given by
\bea
  \log\left(\frac{ g_{(1r)\to(r1)} }{ g_{(11)} }\right)
&=& \int_{-\infty}^\infty \Phi(\theta - \theta_b) 
    L(\epsilon_{r-1}(\theta))
    \frac{\rd\theta}{2\pi}
\;,
\nonumber%\label{eq:gb31}
\\
  \log\left( \frac{g_{(1r)\to((r-1)1)} }{g_{(11)} } \right)
&=& \int_{-\infty}^\infty \Phi(\theta - \theta_b) 
    L(\epsilon_{m+2-r}(\theta))
    \frac{\rd\theta}{2\pi}
\;,
\label{eq:gb32}
\eea
and we propose that the excited state g-functions are given by the
replacement of the ground state TBA functions in these expressions by
the excited state functions: 
\bea
  \log\left(\frac{ g^\sigma_{(1r)\to(r1)} }{ g^\sigma_{(11)} }\right)
&=& \int_{-\infty}^\infty \Phi(\theta - b) 
    L_{r-1}(\epsilon'_{r-1}(\theta))
    \frac{\rd\theta}{2\pi}
\;,
\nonumber%\label{eq:gb41}
\\
  \log\left( \frac{g^\sigma_{(1r)\to((r-1)1)} }{g^\sigma_{(11)} } \right)
&=& \int_{-\infty}^\infty \Phi(\theta - b) 
    L_{m+2-r}(\epsilon'_{m+2-r}(\theta))
    \frac{\rd\theta}{2\pi}
\;,
\label{eq:gb42}
\eea
Using these, in figure \ref{fig:m89plot} we plot
$g_\alpha/g_{(11)}$ 
vs
$b_\alpha^\sigma/b_{(11)}^\sigma$ for the models with $m=2,4$ and $6$.

\newpage

As can be seen, the functions we propose interpolate the conformal
fixed points as required, the $(r1)$ fixed boundary conditions are
local minima of the g-function and the $(1r)$ boundary conditions are
local maxima. 

\subsection{The case of $m$ odd: $M_{2p+3,2p+4}$}

The odd minimal models are slightly different. The ground state TBA
equations are given in \cite{Zamo9b} and the kink ground state
equations are exactly as in \eref{eq:gfn1} but the first excited state TBA
equations do not appear to have been written down so far.
We propose that the kink TBA equations for the case of a critical bulk
are also exactly as in \eref{eq:gfn2}, with the same assignment of
signs. This is asymmetric, for example the signs $s_j$ for
$M_{7,8}$ are $\{+,-,-,+,+,+\}$ nevertheless they appear to be
correct. The boundary state overlaps are also slightly more
complicated having logarithmic terms and cannot be directly deduced 
from the TBA equations using the results of \cite{BLZ94}. We propose
that they are also 
given by \eref{eq:gb32}
and \eref{eq:gb42}; these pass several perturbative and
non-perturbative tests which we plane to report on at greater length
elsewhere. 
We include plots of $g_\alpha/g_{(11)}$ and
$b^\sigma_{\alpha}/b^\sigma_{(11)}$ for $M_{5,6}$ and $M_{7,8}$ in
figure \ref{fig:m89plot}.

\sect{Conclusions}
\label{sec:conc}

We have a global coordinate on a space of boundary flows in the
unitary minimal models in terms of two sets of TBA equations. This
answers the question put in \cite{W2011}, whether such a unified
coordinate can be found. It shows that the space of flows is indeed
compact and smooth.
It would be very interesting to see if the same quantities can be
calculated in the staircase model which was used in \cite{LSS98} and
\cite{DTW10} to study the simple g-function.

Finally we remark that, as has been noted before, the values of
$g_\alpha/g_{(11)}$ and $g^\sigma_\alpha/g^\sigma_{(11)}$ for the
boundary condition $(r,1)$ in 
the model $M_{s,s+1}$ are the same as those for the boundary condition
$(1,r)$ for the model $M_{s-1,s}$ (provided $s>2r$). This can be
explained by the fact that the corresponding conformal defects  are
essentially the same operator -- something we hope to return to later.
\\

\newpage
\noindent{\bf Acknowledgements}
\\

\noindent
I would like to thank P.~Giokas and P.E.~Dorey for very helpful
discussions and STFC grant ST/G000395/1 for support. 
All numerical work was performed using Mathematica \cite{mathematica}.

\label{sec:acc}

\newcommand\arxiv[2]      {\href{http://arXiv.org/abs/#1}{#2}}
\newcommand\doi[2]        {\href{http://dx.doi.org/#1}{#2}}
\newcommand\httpurl[2]    {\href{http://#1}{#2}}

%\newpage

\end{document}